\documentclass{article}

     \usepackage[preprint]{neurips_2022}

\usepackage[utf8]{inputenc} % allow utf-8 input
\usepackage[T1]{fontenc}    % use 8-bit T1 fonts
\usepackage{hyperref}       % hyperlinks
\usepackage{url}            % simple URL typesetting
\usepackage{booktabs}       % professional-quality tables
\usepackage{amsfonts}       % blackboard math symbols
\usepackage{nicefrac}       % compact symbols for 1/2, etc.
\usepackage{microtype}      % microtypography
\usepackage{xcolor}         % colors
\usepackage{url}            % url

\usepackage{mathtools}
\usepackage{cancel}

\usepackage[ruled,linesnumbered]{algorithm2e}
\SetCommentSty{mycommfont}

\usepackage{graphicx}
\usepackage{epstopdf}
\newcommand{\ft}[0]{\footnotesize}
\usepackage{caption}
\usepackage{subcaption}

\usepackage[frozencache=true,cachedir=.]{minted} 
\usepackage{amsthm}
\newtheorem{remark}{Remark}

\title{TPU-KNN\\ K Nearest Neighbor Search at Peak FLOP/s}

\author{%
  Felix Chern \thanks{Equal contributions.} \\
  Google Research \\
  \texttt{fchern@google.com}
  \And
  Blake Hechtman $^*$ \\
  Google Core ML \\
  \texttt{blakehechtman@google.com}
  \And
  Andy Davis $^*$ \\
  Google Core ML \\
  \texttt{andydavis@google.com}
  \And
  Ruiqi Guo \\
  Google Research \\
  \texttt{guorq@google.com} \\
  \And
  David Majnemer \\
  Google Core ML \\
  \texttt{majnemer@google.com}
  \And
  Sanjiv Kumar \\
  Google Research \\
  \texttt{sanjivk@google.com} \\
}

\begin{document}

\maketitle

\begin{abstract}
% This paper presents a nearest neighbor search algorithm achieving TPU (Google Tensor Processing Unit) peak performance, outperforming state-of-the-art GPU algorithms by 4x of similar recalls (R@10 > 0.95). We reached the hardware peak by co-designing the algorithm with an accurate performance model that addresses the memory and instruction bottlenecks quantitatively. Our algorithm comes with an analytical recall guarantee and does not require indexing the database, making it suitable for applications with frequent updates. Our work is available in the open-sourced Jax (0.3.2) and Tensorflow (2.10.0) on Google cloud TPU, and we also provide a standalone repository for the benchmark code.
% 
% See https://docs.google.com/document/d/1FdSVYxFYEhoYYYE1VZu-AjVITXU8AaxEGSc-yb6KJKU/edit for visualized edits.

%This paper presents a novel nearest neighbor search algorithm which can be efficiently implemented on accelerators such as TPU (Google Tensor Processing Unit) to achieve peak utilization, outperforming state-of-the-art GPU implementations with similar level of recall. 

This paper presents a novel nearest neighbor search algorithm achieving TPU (Google Tensor Processing Unit) peak performance, outperforming state-of-the-art GPU algorithms with similar level of recall.
The design of the proposed algorithm is motivated by an accurate accelerator performance model that takes into account both the  memory and instruction bottlenecks. Our algorithm comes with an analytical guarantee of recall in expectation and does not require maintaining sophisticated index data structure or tuning, making it suitable for applications with frequent updates. Our work is available in the open-source package of Jax and Tensorflow on TPU.

\end{abstract}

% High level check list

% \begin{enumerate}
%     \item Baseline 
%     \begin{enumerate}
%         \item \answerTODO{} Do state the high level goal in each section clearly?
%         \begin{itemize}
%             \item \answerYes{} Intro
%             \item \answerYes{} Preliminaries
%             \item \answerYes{} Related work
%             \item \answerYes{} Methodology
%             \item \answerTODO{} Algorithm
%             \item \answerTODO{} Evaluation
%             \item \answerTODO{} Discussion and future work
%         \end{itemize}
%         \item \answerTODO{} Are the symbols consistent?
%         \item \answerTODO{} Describe that we provide a performance model clearly in abstract and in introduction.
%         \item \answerTODO{} Show why the challenge is non-trivial. (not all the utilization are the same)
%         \item \answerTODO{} Use dotted lines in the plots.
%     \end{enumerate}
%     \item Beauty
%     \begin{enumerate}
%         \item \answerTODO{} Novelty: Do we have strong contrast between the old and the new?
%         \item \answerTODO{} Economic: clean formulation of the (performance) problem.
%         \item \answerTODO{} Surprise
%     \end{enumerate}
% \end{enumerate}

\section{Introduction}

%\answerTODO{} Decide to use upper case or lower case for static variables like $N$, $M$, $D$, $K$, etc.

The $K$-nearest neighbor ($K$-NN) search problem has a wide range of applications in machine learning and information retrieval systems, including image search \citep{jiajia2021im2query,babenko2016deep1b},
semantic textual retrieval \citep{liu2009learning,cer2018universal},
anomaly detection \citep{gu2019statistical,omar2013anomaly},
recommendation systems \citep{sarwar2002recommender,zhao2019recommender},
as well as serving as a component for a downstream tasks \citep{borgeaud2021improving, guu2020realm, lindgren2021efficient, shazeer2017outrageously}.
Given a query, the objective of $K$-NN is to identify $K$ closest datapoints from a database of finite number of data points in a vector space. The main challenge of designing a good $K$-NN algorithm is to compute accurate $K$-NN results while being computationally efficient.

%\answerTODO{ Polish sentences}.

%On par with the algorithmic advances on CPU,
Solving the  $K$-NN problem on accelerators has emerging interests from both the academia and the industry \citep{faissgpu,shanbhag2018efficient,songgpu}.
Many accelerators can deliver hundreds of Tera Floating Point Operations Per Seconds (TFLOPS) vital to the neighbor distance computation.
However, utilizing accelerators in $K$-NN problems is not straightforward; multiple issues in data locality, memory bandwidth, and multiple types of hardware parallelism need to be carefully considered to achieve high utilization. 
%Furthermore, modern accelerators have dedicated matrix multiplication units where the operation throughput is different from the other coefficient-wise instructions. Hence, 
% Describe what roofline performance model is.
In this paper we extend the \emph{roofline performance model} \citep{williams2009roofline} to quantify the hardware characteristics accurately.
As a result, we designed a $K$-NN algorithm to reach peak performance by the precise modeling of the accelerators, and our TPU implementation aligned with our predicted performance.

The main contributions of this work are:
\begin{itemize}
\item We extend the roofline model to address the operation throughput differences of the instructions, essential to the algorithm analysis in this paper.
\item We design an approximate $K$-NN algorithm with recall and performance guarantees based on our proposed roofline model.
\item We conduct experiments verifying our TPU implementation of the algorithm accurately aligned with the performance model and achieves state-of-the-art speed-recall trade-offs on standard nearest neighbor search benchmarks.
\end{itemize}

% [[ Let introduce notations in the background?]]

%We limit the scope of our study and benchmarks to single-chip accelerators on million-scale datasets. Billion-scale nearest neighbor search requires either distributed accelerators \citep{faissgpu}, lossy compressions, off-memory search using SSD \citep{chen2021spann,jayaram2019diskann,ren2020hm}, or a combination of the listed techniques, which are exciting but orthogonal research topics to this paper. Nevertheless, readers could easily apply our algorithm to multi-TPU using Tensorflow's \texttt{tf.distribute} \citep{tensorflow2015-whitepaper} or Jax's \texttt{jax.pmap} \citep{jax2018github} programming interfaces. The discussions on multi-TPU is left out because our main argument in this paper is about accurate performance modeling, and the single-chip benchmarks are sufficient to support the study.

\section{Preliminaries}

This section covers the necessary notations to work with the nearest neighbor search problem.
    Given a matrix $\mathbf{A} \in \mathbb{R}^{M\times N}$,
    we let $a_{i,j}$ denote the item at the $i$th row and $j$th column of $\mathbf{A}$, and
    $\mathbf{a}_i$ denote the $i$th \emph{row-vector} of $\mathbf{A}$.
    We use the matrix $\mathbf{X} \in \mathbb{R}^{N\times D}$ to abbreviate a set-representation of a database $\mathbf{X}=\{\mathbf{x}_i\}_{i=1,2,...,N}$ with $N$ data points,
    where each data point $\mathbf{x}_i\in \mathbb{R}^D$ is a row vector of the matrix $\mathbf{X}$ in a $D$ dimensional vector space. The set and matrix representation of database $\mathbf{X}$ are used interchangeably in this paper.

The $K$ nearest neighbor search problem is stated as follows. Given a database $\mathbf{X} \in \mathbb{R}^{N\times D}$
and a query vector $\mathbf{q} \in \mathbb{R}^D$, find the subset $\mathbf{S}^* \subset \mathbf{X}$
collecting the $K$-closest data points to $\mathbf{q}$:

\begin{equation}
    \boxed{\mathbf{S_q}^* = \underset{\mathbf{x}\in\mathbf{X}}{K\text{-argmin }} \mathcal{D}(\mathbf{q}, \mathbf{x}),}
\end{equation}

where $\mathcal{D}(\mathbf{x},\mathbf{y})$ is a distance measure such as Euclidean distance $\mathcal{D}_{\ell^2}(\mathbf{x},\mathbf{y}):=\|\mathbf{x}-\mathbf{y}\|_2$ or the cosine distance $\mathcal{D}_{\cos}(\mathbf{x},\mathbf{y}):=
1-\frac{\langle \mathbf{x}, \mathbf{y} \rangle}{\|\mathbf{x}\|\|\mathbf{y}\|}$. A related problem is the maximum inner product search (MIPS), where the goal is to find the data points
that have the highest inner products with the query:

\begin{equation}
    \boxed{\mathbf{S_q}^* = \underset{\mathbf{x}\in\mathbf{X}}{K\text{-argmax }} \langle \mathbf{q}, \mathbf{x} \rangle.}
\end{equation}

MIPS is equivalent to the cosine similarity search when all data points are $\ell^2$-normalized.
%We use MIPS and cosine $K$-NN interchangeably through out this paper.

\section{Related work}

%\answerTODO{} Moderate statement to compare our methods to others.

Exhaustively searching all pair-wise distances between the query and the entire database is compute-intensive and often infeasible on many platforms. Therefore, a problem extensively discussed in the literature \citep{wang2014hashing,wang2015learning} is to find \emph{approximate nearest neighbors} (ANN) in exchange of speed.
By convention, the quality of ANN is measured by

\begin{equation}
    \boxed{\text{Recall} := \frac{|\mathbf{S_q} \cap \mathbf{S_q}^*|}{|\mathbf{S_q}^*|},}
    \label{eq:recall-def}
\end{equation}

where $\mathbf{S_q}\subset \mathbf{X}$ denotes the set of data points retrieved by the search method.

\paragraph{Compressed domain search} 
\label{ss:ivf} 
One class of ANN approaches is to search on a lossy-compressed problem domain.
These methods are composed in two steps: a) search on compressed representation\footnote{Here we mean data structures like tree, graph, locality sensitive hash etc.} of the original problem to find a set of candidate data points, b) compute the distances between the query and the candidate data points to select the top-$K$ results. Since only a subset of data points requires the exact distance computation, the overall cost is reduced.

The two steps can be composed in arbitrary ways. Locality sensitive hashing \citep{andoni2015practical,neyshabur2015symmetric} applies search followed by scoring;
tree-search \citep{muja2014scalable, dasgupta2008random} applies the two steps recursively; graph-search \citep{malkov2018efficient} iterates between two steps until the stopping condition is met.
And the inverted file (IVF) method \citep{jegou2010product, babenko2014inverted, baranchuk2018revisiting, guo2020accelerating} search on subset of data points indexed by the k-means centroids.

We see that there are two major challenges with the compressed domain search:

\begin{itemize}
    \item Fractional search has a poor cache reuse rate because the candidate data points for each query rarely overlaps. We show optimizing the cache usage has a huge headroom for accelerators in Section \ref{ss:mem-bound}.
    \item Tweaking the speed-recall trade-off is data-dependent and non-trivial to tune. The key result of \cite{beyer1999nearest} states that the distance contrast of neighbors diminishes with increasing dimensionality (also known as the curse of high dimensionality). Furthermore, the key result of \cite{rubinstein2018hardness} states that sub-linear time nearest neighbor search with high recall is impossible for Euclidean, Manhattan, or Hamming distance; otherwise, it contradicts the Strong Exponential Time Hypothesis \citep{impagliazzo1999complexity}.
\end{itemize}

%\answerTODO{} The following sentence requires polishing.

Our work takes an opposite approach to focus on machine efficiency with zero search space pruning.  Moreover, since our method computes all the distances, it is immune to the curse of high dimensionality.

\paragraph{Accelerators}
In this paper, the phrase \emph{accelerators} represents a class of specialized hardware to accelerate machine learning workloads. In particular, we are interested in the novel platforms that deliver high FLOP/s for distance computation, namely
Google TPU V3, V4, Nvidia GPU V100, and A100 in our analysis and evaluation.

Modern accelerators have special computation units for matrix multiplication, providing a higher operation throughput over the regular coefficient-wise operations. The corresponding units are tensor cores in Nvidia GPUs \citep{markidis2018nvidia} and systolic arrays in Google TPUs \citep{jouppi2017datacenter,norrie2021design}. Addressing these operation throughput differences is essential to our algorithm design.

%State-of-the-art top-$K$ aggregation method \citep{blum1973time} achieves $\mathcal{O}(MN)$ amortized complexity on CPU.
While accelerators excel in parallelism, developing an efficient $K$-selection algorithm on accelerators is still an active research area \citep{monroe2011randomized,shanbhag2018efficient,faissgpu,songgpu}. Accelerators with higher FLOP/s introduce a higher opportunity cost of computing the $K$-selection problem instead of the distance computation. The trend of the increasing FLOP/s in accelerators motivated us to optimize the FLOP/s usage by reducing the time required for computing $K$-selection. 

%\answerTODO{} Our method uses an exhaustive search followed by an approximate top-$K$ aggregation to minimize the opportunity cost of FLOP/s.

\section{Methodology}
\label{s:methodology}

This section presents a performance model to identify non-trivial bottlenecks on multiple platforms and demonstrates some fundamental limits when designing algorithms for $K$-NN and related problems, and we see that the cache inefficiency of the compressed domain methods introduces a significant cost on accelerators.

We model the accelerator's runtime as executing a sequence of \emph{computation kernels}, where each kernel is a compiled subroutine on the accelerator used by the main program on the CPU. A kernel may be composed of one or several high-level operators: Einsum, ReLU, ArgMax, etc., and each kernel can have different performance characteristics.

Given a sequence of kernels $k_i$, we let $W_i$ denotes the total amount of work and $P_i$ denotes the operational speed. Our goal is to estimate the total time of a program:

\begin{equation}
    t = \sum_i\frac{W_i}{P_i}.
\end{equation}

In the following example, we focus on the MIPS problem. Let $\mathbf{Q} \in \mathbb{R}^{M\times D}$ and
$\mathbf{X}\in \mathbb{R}^{N\times D}$ denote the queries and the database, the runtime of a generic approximate-MIPS program can be modeled as

\begin{equation}
    \boxed{
    t = \frac{\lambda W_{\mathcal{D}}}{P} + \mathcal{O}(\text{Auxiliary})
      \ge \frac{\lambda W_{\mathcal{D}}}{P}, % = \frac{2\lambda NMD}{P},
    }
    \label{eq:runtime}
\end{equation}

where $W_{\mathcal{D}}$ denotes the total FLOPs required for searching the entire database, and $\lambda$ denotes the \emph{search fraction}. We note that $P$ varies by algorithm and platform. Traditionally, compressed domain search methods minimize $\lambda$ but sacrifice cache efficiency. Our method use an alternative route to optimize $P$ instead.

%We follow the FLOPs calculation used in standard matrix multiplications where multiply and add are treated as two operations.

%The compressed domain search methods focus on minimizing $\lambda$. Our work has $\lambda=1$ but we optimize $P$ to machine's peak performance. We show improving $P$ has a huge headroom in the following sections.

\subsection{Instruction throughput-aware roofline model}
This subsection describes how we model the kernel-dependent performance $P$ on multiple platforms with a small extension of the roofline model.

The \emph{classic roofline model} \citep{williams2009roofline} is a function of machine peak performance $\pi$ measured in FLOP/s, machine peak memory bandwidth $\beta$ measured in bytes/s, and arithmetic intensity $I_{\text{MEM}}$ expressed as the ratio of floating-point operations performed to data movement (FLOP/byte). The model states the performance is bounded by $P \le \min(\pi, \beta \times I_\text{MEM})$.

We desire to model kernels that has a mixture of floating point operations accelerated by dedicated hardware as well as other coefficient-wise operations. The coefficient-wise operations are abbreviated as COPs. Almost every non matrix multiplication operations are COPs, including vectorized add, multiply, compare, conditional-move, etc. We use the symbol $\gamma$ for peak COP/s on platforms, and define the instruction throughput intensity $I_\text{COP}$ as the ratio between the number FLOPs and the number of COPs performed in a kernel (FLOP/COP). The attainable performance of a kernel is bounded by:

\begin{equation}
\boxed{
    P \le \min
    \begin{cases}
    \pi \\
    \beta \times I_\text{MEM} \\
    \gamma \times I_\text{COP} .
    \end{cases}
}
    \label{eq:roofline}
\end{equation}

The statement is self-explanatory because the inadequate resources impede the kernel throughput. Table \ref{t:roofline} lists the properties of selected accelerators for our analysis\footnote{Readers can find these numbers from the accelerators' specification sheets.}. The roofline model is commonly used in accelerator profiling tools but not as frequently discussed in algorithm designs. The following sections show how the model prevents pitfalls due to the hardware constraints.

\begin{table}
  \caption{Hardware specifications for the generalized roofline model}
  \label{t:roofline}
  \centering
  \begin{tabular}{lrrr}
    \toprule
    %\multicolumn{2}{c}{Part}                   \\
    %\cmidrule(r){1-2}
    Name     & $\pi$ (TFLOP/s) & $\beta$ (GB/s) & $\gamma$ (TCOP/s) \\
    \midrule
    GPU V100 & 125 & 900 & 15.7 \\
    GPU A100 & 312 & 1555 & 19.5 \\
    TPU V3 & 126 & 858 & 4.0 \\
    TPU V4 & 274 & 1144 & 4.3 \\
    \bottomrule
  \end{tabular}
\end{table}

\subsection{The memory bandwidth bound}
\label{ss:mem-bound}

This subsection demonstrates how to evaluate if a kernel hits the memory bandwidth wall. We associate the distance computation with three levels of BLAS \citep{dongarra1990set}. Level 1 BLAS describes vector operations on non-consecutive memory access, such as computing distances while traversing through a graph. Level 2 BLAS represents scoring a query with consecutively stored data points. Level 3 BLAS expresses batched query-database distance computation, often used in brute-force scoring.

%\answerTODO{} How do we highlight the statement? Should we wrap it into a theorem?

Compressed domain searches are either level 1 or 2 BLAS due to the cache inefficiency. It has $\mathcal{O}(1)$ memory arithmetic intensity because the number of FLOPs is proportion to the bytes read. Combining \eqref{eq:runtime} and \eqref{eq:roofline} we have the following remark:
 
\begin{remark}
\label{rm:compressed-search}
Distance computations in compressed domain searches are memory bandwidth bounded. In our model, the runtime is lower bounded by: $t \ge \mathcal{O}\left(\lambda W_\mathcal{D}/\beta\right)$.
\end{remark}

To estimate the memory arithmetic intensity for level 3 BLAS,
we continue to use $\mathbf{Q} \in \mathbb{R}^{M\times D}$ and
$\mathbf{X}\in \mathbb{R}^{N\times D}$ for denoting queries and database. In many $K$-NN applications $N$ and $M$ are much greater than $D$. The corresponding memory arithmetic intensity is:

\begin{equation}
    I_\text{MEM} = \frac{2MND}{4MN + o(MN)} \approx \frac{D}{2}.
    \label{eq:blas3}
\end{equation}

The largest term in the denominator of \eqref{eq:blas3} is the $4MN$ bytes of the query-database distances. We omit the insignificant terms and refer readers to \cite[][Section 1.5.4]{golub2013matrix} for a comprehensive review on memory transfers in block matrix multiplications. 

\begin{figure}
\input{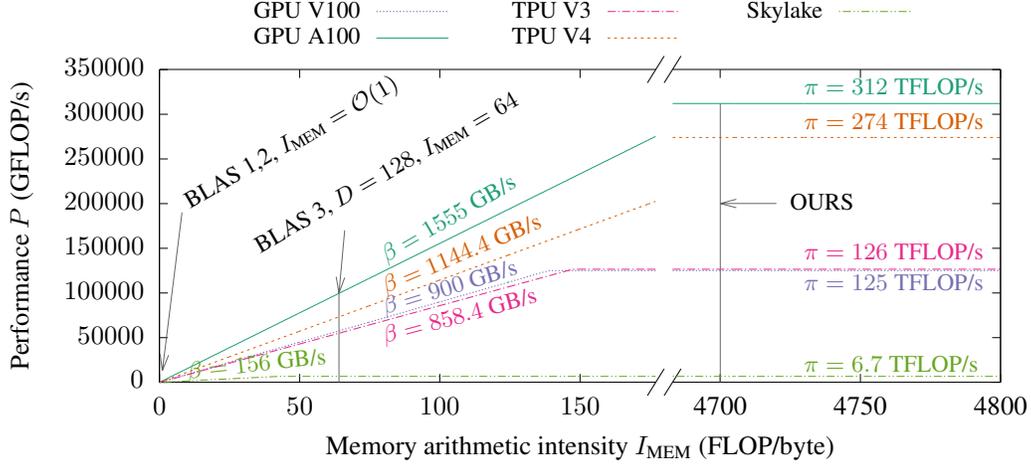}
\caption{
Memory rooflines of accelerators and a dual-sockets Intel skylake machine as a baseline. Each colored line denotes the maximum performance a platform could achieve, and each vertical line represents the memory arithmetic intensity of an algorithm. The intersections of the lines show the maximum performance of an algorithm could achieve on a platform. We label three levels of BLAS kernels and our algorithm described in Section \ref{s:algorithm}.
}
\label{fig:blasroofline}
\end{figure}

Figure \ref{fig:blasroofline} shows that the distance scoring kernels of different BLAS levels can easily hit the memory bandwidth wall. In order to attain high performance, we designed our algorithm to aggregate the results within the kernel to avoid writing the $\mathcal{O}(MN)$ bytes into memory.

\subsection{The instruction bandwidth bound}
\label{ss:ins-roofline}

The use of COPs (non matrix multiplication instructions) introduce another slowdown. We let $C$ denotes the number of COPs used per dot-product score in a kernel equipped with COPs and matrix multiplication instructions. There are $M\times N$ dot-product scores, so the total COPs used in a kernel is $CMN$. To prevent hitting the COPs bandwidth wall, we must satisfy:

\begin{align}
    I_\text{COP} &= \frac{2\cancel{MN}D}{C\cancel{MN}} \ge \frac{\pi}{\gamma}, \\
    \Rightarrow C & \le \frac{2D \times \gamma}{\pi}.
    \label{eq:c-limit}
\end{align}

The number of COPs we can afford in the kernels is scarce. We take $D=128$ as an example and substitute it into \eqref{eq:c-limit}. We can only use 4 coefficient-wise instructions per dot-product for TPU V4, and 16 for GPU A100. We conclude with the following remark:

\begin{remark}
\label{rm:exact-k}
Exact and generic $K$-selection algorithm cannot be efficiently implemented with the coefficient-wise operations for the selected platforms (GPU V100, A100, TPU V3 and V4).
\end{remark}

Because of Remark \ref{rm:exact-k}, we develop an approximate approach to achieve the peak performances.

%This headroom analysis motivates our two kernels algorithm design: the first kernel computes the distance and partially aggregate the output with minimal COPs used, and the second kernel computes regular $K$-selection on the reduced problem. See the next section for the full description.

\section{Algorithm}
\label{s:algorithm}
\begin{algorithm}
\caption{PartialReduce for MIPS}\label{alg:approx:mips}
\KwIn{$\mathbf{Q}\in \mathbb{R}^{M\times D}$ Batch queries}
\KwIn{$\mathbf{X}\in \mathbb{R}^{N\times D}$ Database}
\KwIn{$2^W$ Bin size}
\KwOut{$\mathbf{V} \in \mathbb{R}^{M\times L}$ Top-$K$ values}
\KwOut{$\mathbf{A} \in \mathbb{N}^{M\times L}$ Top-$K$ indices}
\SetKwFunction{ShiftRight}{ShiftRight}

\BlankLine
\For{$i \gets 1$ \KwTo $M$}{
  \For{$j \gets 1$ \KwTo $N$}{
    $y_{i,j} \gets \langle\mathbf{q}_i, \mathbf{x}_j \rangle$ \;
    $l \gets \ShiftRight{j, W}$ \tcc*[r]{Unrolled and does not cost COP}
    $b \gets y_{i,j} > v_{i,l}$ \tcc*{COP 1: Vectorized compare}
    $v_{i,l} \gets \textbf{ if } b \textbf{ then } y_{i,j}  \textbf{ else } v_{i,l}$
          \tcc*{COP 2: Vectorized conditional move}
    $a_{i,l} \gets \textbf{ if } b \textbf{ then } j \textbf{ else } a_{i,l}$
          \tcc*{COP 3: Vectorized conditional move}
  }
}
\end{algorithm}

Our algorithm consists of two kernels:

\begin{enumerate}
    \item PartialReduce kernel computes the distances and partially aggregate the results from $M\times N$ distances to $M\times L$ distances with original indices.
    %The distance aggregation happens in registers for performance and prevents the potential out-of-memory error.
    \item ExactRescoring kernel is an \emph{optional} kernel that aggregates the final top-$K$ results. The complexity is $\mathcal{O}(ML\log^2(L))$ by a bitonic sort followed by a truncation.
\end{enumerate}

The PartialReduce kernel is where most of the time and compute takes place. See Algorithm \ref{alg:approx:mips} for an outline of the algorithm.
We collect top-1 distances from the $L$ non-overlapping bins of size $2^W$ for each query, resulting high arithmetic intensities:

\begin{align}
    I_\text{MEM} &\approx \mathcal{O}\left(\min \left(M, N\right)\right) \label{eq:pr:imem},\\
    I_\text{COP} &= \frac{2\cancel{MN}D}{C\cancel{MN}} = \frac{2D}{C} \label{eq:pr:icop}.
\end{align}

We show these arithmetic intensities can achieve high performance on real world database in section \ref{ss:eval-peak}. See Appendix \ref{ap:detail} for the detailed expansion of the algorithm and how the arithmetic intensities are derived.

\subsection{Recall estimation}
\label{ss:recal-est}

This section shows the PartialReduce kernel can achieve high recall with good speed.
We reformulate our problem in terms of balls and bins. We have $K$ balls representing the top-$K$ distances that are thrown into $L$ bins. The location of each ball is chosen independently and uniformly at random.
We let $\mathbf{Z}$ denotes the random variable of the number of balls that does not share the bin with other balls. Following the recall definition \eqref{eq:recall-def} we have:

\begin{equation}
    \text{Recall} = \frac{\mathbf{Z}}{K},
\end{equation}

which is a standard Birthday problem:

\begin{equation}
\mathbb{E}[\text{Recall}] = \frac{\mathbb{E}[\mathbf{Z}]}{K} = \left(\frac{L-1}{L}\right)^{K-1}.
\label{eq:recall}
\end{equation}

Our goal is to find the minimal $L$ such that the expected recall is greater equals to 
the target recall $r$.
Finding $L$ is simple because \eqref{eq:recall} is invertible in the natural range $0<r<1$.

\begin{equation}
\boxed{
    \mathbb{E}[\text{Recall}] \ge r
   \Rightarrow L \ge \frac{1}{1-r^{1/(K-1)}} 
                 \approx \frac{K-1}{1-r}.}
    \label{eq:l:approx}
    %\approx \frac{1}{1-\exp[\frac{r-1}{K-1}]}.
\end{equation}

The approximation in \eqref{eq:l:approx} follows from Appendix \ref{ss:recall-approx}.
Since $L$ is at the order of $K$, and in most applications $K\ll N$, the cost of the ExactRescoring kernel is amortized out. Thus we affirm the claim that our method attains high performance with an analytical recall guarantee. \qed

% Our implementation derives the number of bins $L$ and the window size $2^W$ from user specified recall target at compile time.
% The number of bins $L$ only depends on $K$ and the recall target, which makes the PartialReduce kernel easy to achieve high performance with a large enough $N$ and $M$, affirming our claim.

\section{Evaluation}
\label{s:eval}

In this section, we show that our proposed algorithm and implementation is near the hardware limit and leads to superior performance over the baselines of similar recalls.
We applied our algorithm to two datasets from the public ANN benchmarks \citep{aumuller2020ann}.
In our first evaluation, we compares the measured FLOP/s to the theoretical peak governed by the proposed refinement of the roofline model \eqref{eq:roofline}, proclaiming our implementation is reaching the hardware peak performance.
In the second benchmark, we compare the end-to-end performance with competitive baselines with pre-tuned parameters. We plot each algorithm's speed-recall curve and show ours achieves the state-of-the-art.

\subsection{Comparison with the theoretical peak}
\label{ss:eval-peak}

\begin{figure}
\include{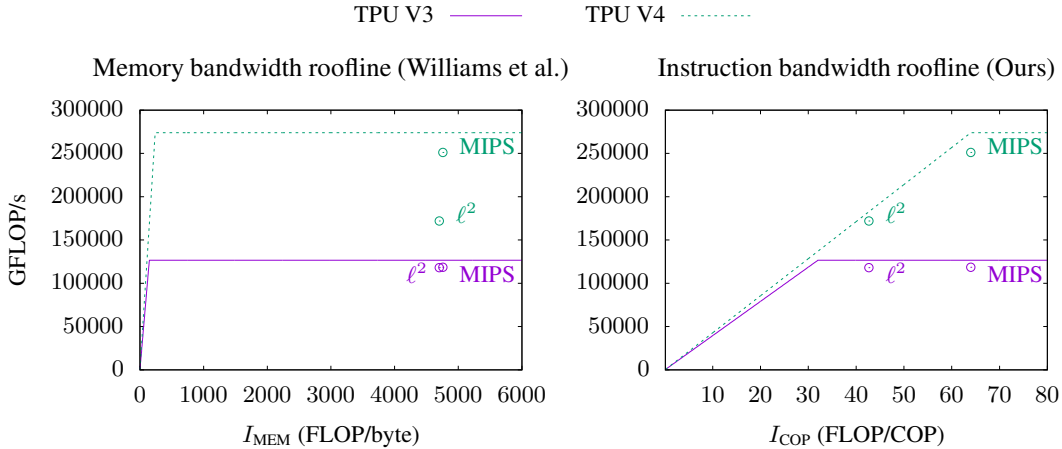}
\caption{Roofline plots for MIPS and $\ell^2$ search benchmarks using the PartialReduce kernel. The colored lines denotes the attainable performance derived from Table \ref{t:roofline}. The figure on the left shows none of the benchmark is memory bandwidth limited. The figure on the right shows that our model gives a much tighter bound for $\ell^2$ on TPU V4. See also Appendix \ref{ap:benchmark} for detailed deviation of the numbers.}
\label{fig:multi_roofline}
\end{figure}

This section shows that our refined roofline model \eqref{eq:roofline} captures additional performance characteristic over the classic roofline model, and demonstrates our kernels are having near optimal performances.
We select the Glove\footnote{Released in Apache license 2.0.} \citep{pennington2014glove} and Sift\footnote{Released in CC0 public domain.} \citep{jegou2010product} datasets from the ANN benchmarks. Their corresponding distances are the cosine distance and the Euclidean distance. See the code snippets in Appendix \ref{ap:mips} and \ref{ap:l2}.

See Figure \ref{fig:multi_roofline}, the colored lines represent machines' max performances, and the dots represent each benchmark with its measured FLOP/s. The classic roofline on the left shows that our in-cache aggregation strategy has a large memory arithmetic intensity ($\sim$4,700) exceeding the memory bandwidth ridge points $\pi/\beta$. However, it is difficult to diagnose why the Euclidean distance search does not perform well on TPU V4 from the classic roofline plot.

Fortunately, when combined with the instruction bandwidth roofline we can tell the performance regression is caused by hitting the coefficient-wise operation throughput wall. Therefore we affirms the claim that our MIPS solution is reaching the peak FLOP/s, and our Euclidean distance search solution is meeting the compute bound on TPU V4 and attaining the peak FLOP/s on TPU V3.

\subsection{Recall-speed benchmark}

To evaluate the effectiveness of the $K$-NN algorithm in a realistic setting, we adopted the methodology of public ANN benchmarks \citep{aumuller2020ann} to compare the end-to-end performance against other methods.
The typical ANN benchmarks are only performed on a single platform. However, it is non-trivial to either port our TPU algorithm to GPU or vice versa. Alternatively, we selected the following GPUs with parity in peak performance to TPU (Table \ref{t:roofline}).

We select the Faiss GPU \citep{faissgpu} implementation as our baseline. Faiss provides three algorithms: Flat, IVF-flat, and IVF-PQ. The Flat algorithm performs a brute-force search, and the IVF-Flat and IVF-PQ algorithms corresponds to the inverted file method with and without the product quantization \citep{jegou2010product, faissgpu}. 
We use the repository's suggested inverted file size (16384) in the IVF methods.

\begin{figure}
    \begin{subfigure}[b]{\textwidth}
    \include{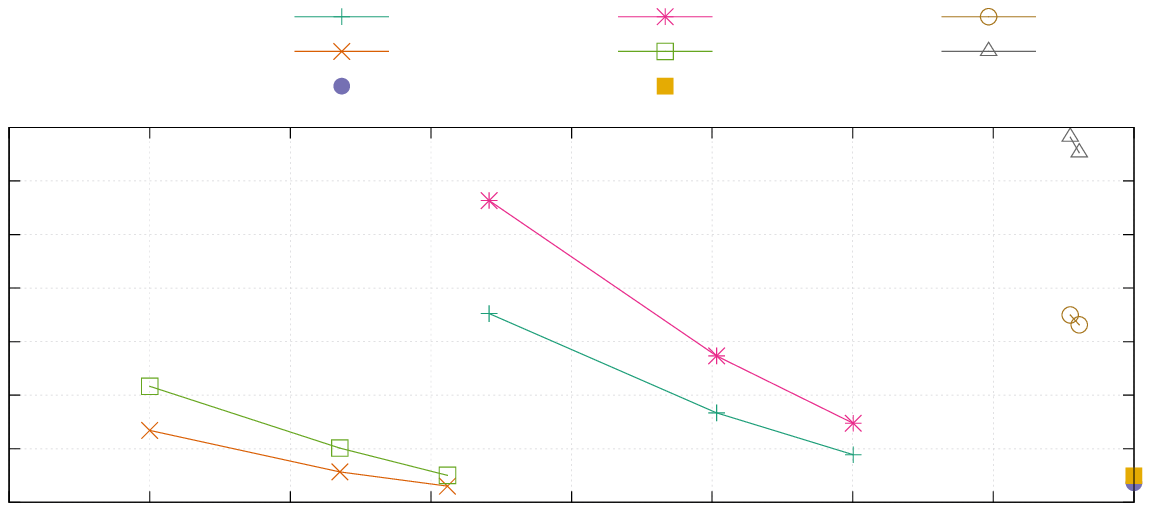}
    %\includegraphics{glove}
    %\caption{Speed-recall trade-off on GloVe1.2M with recall 10@10}
    \label{fig:glove}
    \end{subfigure}
    \begin{subfigure}[b]{\textwidth}
    \include{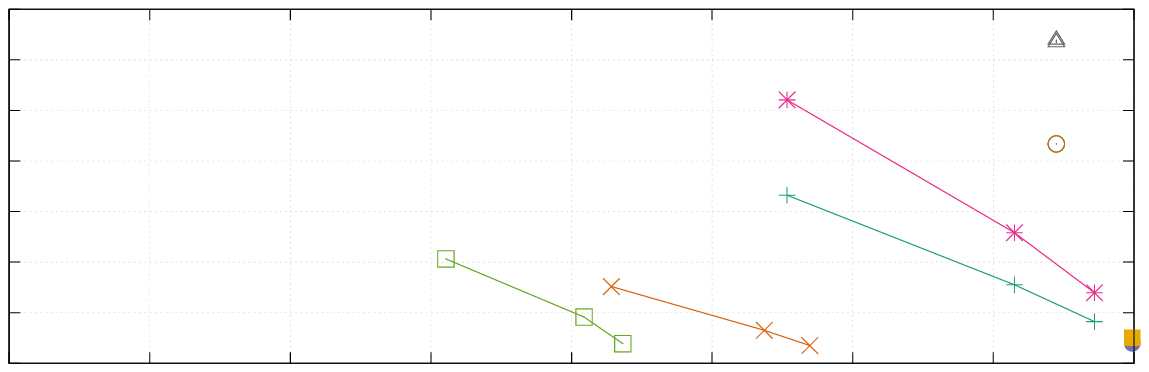}
    %\includegraphics{sift}
    %\caption{Speed-recall trade-off on SIFT1M with recall 10@10}
    \label{fig:sift}
    \end{subfigure}
    
    \caption{Speed-recall trade-off on Glove1.2M and Sift1M. Up and to the right the better. The GPU methods (ivf-flat, ivf-pq, and flat) are released by Faiss \citep{faissgpu}. For each ivf$^*$ benchmark, the search fractions are $\lambda=\{0.24\%, 0.61\%, 1.22\%\}$. We note that the recall differences between datasets with similar ivf search configurations is a known problem asserted by \cite{rubinstein2018hardness}.}
    \label{fig:e2e}
\end{figure}

Figure \ref{fig:e2e} shows our performance significantly outperforms competing methods in the high recall regions. We highlight that our method has a consistent recall-speed trade-off over different datasets,
because our recall only rely on the order statistics instead of the information encoded in the compression domain search methods, which may vary by the datasets. Since our method scores all the pair-wise distances, our method is immune from the curse of high dimensionality.

\section{Discussion and future work}

We limit our experiments and discussion to single-chip accelerator $K$-NN performance of dense vectors. Accelerators performance on sparse vectors follow a completely different paradigm due to random memory access.
Our algorithm can be naturally extended to multi-chip or distributed settings to handle billion scale datasets through Tensorflow's \texttt{tf.distribute} \citep{tensorflow2015-whitepaper} or Jax's \texttt{jax.pmap} \citep{jax2018github} programming interfaces.

It is also possible to use our operations in conjunction with other strategies, including dimension reduction, quantization and tree search \citep{van2009dimensionality,jegou2010product,wang2014hashing},
because many compressed domain search methods use brute-force distance computation on its auxiliary data structures before performing the fractional search.
We note that heterogeneous architectures with off-HBM storage such as host-RAM or even SSD \citep{chen2021spann,jayaram2019diskann,ren2020hm} are great starting points for future research.

Finally, while our refinement of the roofline model handles kernel with mixture of instructions that have different throughput, there are many studies that have extended the roofline model to related topics in recent years: GPU warp instruction roofline \citep{ding2019instruction}, time-based roofline \citep{wang2020time}, roofline for multiple cache tiers \citep{yang2021hierarchical}, and energy rooflines \citep{choi2013roofline,lopes2017exploring}.
Readers may refer to these models for some analysis that are left out, such as the auxiliary work in \eqref{eq:runtime}.

\section{Conclusion}

Accelerator-based machine learning has become the mainstream in academics and industries. However, the performance characteristics of accelerators are counter-intuitive and difficult to program. In this paper, we propose a roofline-based complexity analysis framework to discuss the optimality of the algorithms without low-level optimization details: unrolling factors, batch window sizes, vectorization, and systolic array scheduling, which are platform-dependent and lengthy to read. We demonstrated several examples of inferring the hardware performance limits by simply addressing the kernel's total FLOPs, byte transferred, and the number of coefficient-wise instructions used.  Our refined model foreshadowed non-trivial performance regression caused by the coefficient-wise instructions bandwidth. We took it into account to design a new algorithm for $K$-NN and achieved peak performance on TPU. Finally, our experiments showed that our method outperformed state-of-the-art baselines on platforms with similar performance characteristics, which are known to be hard to beat.

\begin{ack}

We would like to thank the XLA team for the continuous effort on developing
the state-of-the-art compiler and the full support on enabling our new op: \texttt{approx\_max\_k}.
We are also grateful to the Google ScaNN team for the joint effort on bridging the impactful
$K$-NN problem into the accelerator ecosystem.
Last but not least, we thank to Peter Hawkins, Edward Schwartz, and Mani Varadarajan for code reviews in Jax and Tensorflow.

This work was performed and funded by Google.

% maybe Hagen Wierstorf for gnuplot styles in gnuplotting.org.

\end{ack}

\bibliography{ref}
\bibliographystyle{apalike}

\newpage

\appendix

\section{Appendix}

\subsection{MIPS implementation}
\label{ap:mips}

\begin{listing}[h]
\begin{minted}[frame=lines]{python}
import jax
@jax.jit
def MIPS(query, database):
  scores = jax.numpy.einsum('ik,jk->ij', query, database)
  return jax.lax.approx_max_k(scores, k=10, recall_target=0.95)
\end{minted}
\caption{Jax code for maximum inner product search (MIPS)}
\label{mips:jax}
\end{listing}

Listing \ref{mips:jax} demonstrates a maximum inner product search (MIPS) kernel implemented with Jax. Tensorflow users can use the \texttt{tf.math.approx\_max\_k} interface; the underlying XLA compiler delivers the same kernel.
There are several options to control the behavior of \texttt{approx\_max\_k}, listed as below:

\begin{enumerate}
    \item \texttt{reduction\_dimension} specifies the dimension in which to search. Default -1 (the last dimension.)
    \item \texttt{recall\_target} derives the number of bins $L$ of the PartialReduce kernel output. Default 0.95.
    \item \texttt{reduction\_input\_size\_override}. When set to a positive value, it overrides the size determined by input for evaluating the recall and bin numbers $L$. Users could use this option to control the kernel output size in the distributed environment.
    \item \texttt{aggregate\_to\_topk}. When set to \texttt{True} emits the ExactRescoring kernel. Default: \texttt{True}.
\end{enumerate}

We also provid a separated \texttt{approx\_min\_k} interface for finding minimum distances, which is used in the Euclidean distance search.

\subsection{Euclidean distance search implementation}
\label{ap:l2}

\begin{listing}[h]
\begin{minted}[frame=lines]{python}
@jax.jit
def l2nns(qy, db, db_half_norm):
  dots = jax.numpy.einsum('ik,jk->ij', qy, db)
  dists = db_half_norm - dots
  return jax.lax.approx_min_k(dists, k=10, recall_target=0.95)
\end{minted}
\caption{Jax code for nearest neighbor search in the Euclidean space.}
\label{l2nns}
\end{listing}

Listing \ref{l2nns} is the Jax implementation of Euclidean space nearest neighbor search. We made a few adjustments to speed up the computation. First, for every query vector $\mathbf{q}$, the following search produces the same result:

\begin{align}
     \mathbf{S}_{\ell^2}^* &= \underset{\mathbf{x}\in\mathbf{X}}{K\text{-argmin }} \|\mathbf{q} - \mathbf{x}\|_2 \\
                  &= \underset{\mathbf{x}\in\mathbf{X}}{K\text{-argmin }} \|\mathbf{q} - \mathbf{x}\|^2 \\
                  &= \underset{\mathbf{x}\in\mathbf{X}}{K\text{-argmin }}
                    \|\mathbf{q}\|^2 + \|\mathbf{x}\|^2 - 2\langle \mathbf{q}, \mathbf{x} \rangle \\
                  &= \underset{\mathbf{x}\in\mathbf{X}}{K\text{-argmin }}  \|\mathbf{x}\|^2 - 2\langle \mathbf{q}, \mathbf{x} \rangle
                  \label{eq:l2-suboptimal}
\end{align}

The last equation holds because omitting the query norm does not affect the rank of each result. Nevertheless, \eqref{eq:l2-suboptimal} still uses 2 COPs for the distance computation (one subtract and one multiplication). We can further reduce it to 1 COP by pre-computing the halved norm:

\begin{equation}
    \mathbf{S}^*_{\ell^2} = \underset{\mathbf{x}\in\mathbf{X}}{K\text{-argmin }}  \frac{\|\mathbf{x}\|^2}{2} - \langle \mathbf{q}, \mathbf{x} \rangle \label{eq:l2final}
\end{equation}

\subsection{MIPS PartialReduce kernel internals}
\label{ap:detail}

The MIPS PartialReduce kernel follows the standard numerical computation best practices to utilize the cache usage with the \emph{temporal} and \emph{spatial locality}. See Algorithm \ref{alg:approx:mips:detail} that uncovers the omitted details in Algorithm \ref{alg:approx:mips}.

\begin{algorithm}
\caption{Detailed PartialReduce kernel for MIPS}
\SetKw{KwStep}{step}
\SetKwFunction{RegisterAlignedShiftRight}{RegisterAlignedShiftRight}

\label{alg:approx:mips:detail}
\KwIn{$\mathbf{Q}\in \mathbb{R}^{M\times D}$ Batch queries}
\KwIn{$\mathbf{X}\in \mathbb{R}^{N\times D}$ Database}
\KwIn{$2^W$ Bin size}
\KwOut{$\mathbf{V} \in \mathbb{R}^{M\times L}$ Top-$K$ values}
\KwOut{$\mathbf{A} \in \mathbb{N}^{M\times L}$ Top-$K$ indices}

\tcc{Block iteration over rows}
\For{$ii \gets 1$ \KwTo $M$ \KwStep $ib$ \label{alg:mips:ib}}{
\tcc{Block iteration over columns}
  \For{$jj \gets 1$ \KwTo $N$ \KwStep $jb$ \label{alg:mips:jb}}{
  \tcc{$i, j, k$ and $l$ are often unrolled or even vectorized}
    \For{$i \gets ii$ \KwTo $ii+ib-1$ \label{alg:mips:inner}}{
      \tcc{Starts the inner loop of the systolic arrays}
      $\mathbf{y}_i \gets \mathbf{0}$ \;
      \For{$k \gets 1$ \KwTo $D$}{
        $m\gets q_{i,k}$\;
        \tcc{Vectorized FMA (fused-multiply-add)}
        \For{$j \gets jj$ \KwTo $jj+jb-1$}{
          $y_{i,j} \gets y_{i,j} + m \cdot x_{j,k}$ \;
        }
      }
      \tcc{Ends the inner loop of the systolic arrays}
      \For{$j \gets jj$ \KwTo $jj+jb-1$}{
        \tcc{The exact $j\rightarrow l$ mapping is determined by the compiler backend}
        $l \gets \RegisterAlignedShiftRight{j, W}$ \;
        $b \gets y_{i,j} > v_{i,l}$ \tcc*{COP 1: Vectorized compare}
        $v_{i,l} \gets \textbf{ if } b \textbf{ then } y_{i,j}  \textbf{ else } v_{i,l}$
          \tcc*{COP 2: Vectorized conditional move}
        $a_{i,l} \gets \textbf{ if } b \textbf{ then } j \textbf{ else } a_{i,l}$
          \tcc*{COP 3: Vectorized conditional move}
      }
    }
  }
}
\end{algorithm}

The \emph{temporal locality} refers to reusing previously accessed items. In line \ref{alg:mips:ib}, we iterate by blocks of queries. The block of queries is reused in the inner loops, achieving the temporal locality.

The \emph{spatial locality} refers to accessing items nearby previously accessed. The block iteration loads a chunk of data points (line \ref{alg:mips:jb}) to achieve this optimization. The same block iteration structure may apply recursively for multiple cache hierarchies till the register level.

The inner loops (indexed by $i$, $j$, and $k$ in line \ref{alg:mips:inner}) should be unrolled or even vectorized so that every cycle can produce multiple results via the SIMD (Single Instruction Multiple Data) instructions or systolic arrays.

The algorithm principle is the same on every platform, except that the block factor and vectorization sizes are platform-dependent. We refer readers to \citep{golub2013matrix} for more details.

\paragraph{Estimate memory transfers}

In Algorithm \ref{alg:approx:mips:detail}, memory transfer for each portion of the data is listed below:

\begin{itemize}
    \item Query is only transferred once. Takes $4MD$ bytes.
    \item Database is transferred $\frac{M}{ib}$ times. Takes $4ND\frac{M}{ib}$ bytes.
    \item Outputs are transferred once. Takes $2\times 4 ML$ bytes.
\end{itemize}

The precise formulation for memory arithmetic intensity is

\begin{equation}
    I_\text{MEM} = \frac{2MND}{4\left(MD + \frac{MND}{ib} + 2ML \right)},
\end{equation}

which would approach $\mathcal{O}(\min(M, N))$ as long as $L\ll \min(M, N)$ and the compiler chooses a large enough $ib$ to minimize the database transfer.

\paragraph{Estimate COPs used}

The PartialReduce kernel listed in Algorithm \ref{alg:approx:mips} and \ref{alg:approx:mips:detail} only use $C=3$ per dot-product. However, there are two cases that would increase the number of COPS on TPU due to the implementation constraints:
 \begin{enumerate}
     \item When the dimension $D$ is not multiple of 128, $C$ increases by 1.
     \item When the database size $N$ is not power-of-2, $C$ increases by 1.
 \end{enumerate}

See Appendix \ref{ap:benchmark} on how it affects the real world benchmarks.
% Minimizing the number of COPs per dot-product is crucial for us to attain high performance. Therefore, we must address the number of COPs used accurately in our algorithm design and analysis.

\subsection{Lower bound approximation of the number of bins}
\label{ss:recall-approx}

We care about the number of bins $L$ in the high recall region. Let the target recall $r=1-\epsilon$, we have

\begin{align}
    L &\ge \frac{1}{1-r^{1/(K-1)}} \\
      &= \frac{1}{1-(1-\epsilon)^{1/(K-1)}}  \\
      &\approx \frac{1}{1-\exp[\frac{\epsilon}{K-1}]} \label{eq:e:approx} \\
      &= \frac{1}{1-(1-\frac{\epsilon}{K-1} + o(\epsilon))} \label{eq:taylor} \\
      &\approx \frac{K-1}{\epsilon}.
\end{align}

The approximation in \eqref{eq:e:approx} follows from $(1-\epsilon)^a = (1-\epsilon)^{\frac{1}{-\epsilon}(-\epsilon a)} \rightarrow e^{-\epsilon a}$ as $\epsilon \rightarrow 0$, and \eqref{eq:taylor} follows from the Taylor expansion. \qed

\subsection{Benchmark details}
\label{ap:benchmark}

\begin{table}
  \caption{Dataset properties and the benchmark results}
  \label{t:dataset}
  \centering
  \begin{tabular}{lrr}
    \toprule
        &  Glove1.2M & Sift1M \\
    \midrule
    Dimension $D$ & 100 (Padded to 128) & 128 \\
    Database size $N$ & 1,183,514 & 1,000,000\\
    Query size $M$ & 10,000 & 10,000 \\
    Distance & Cosine & Euclidean \\
    $C$ & 4 & 6 \\
    $I_\text{MEM}$ & 4,758 & 4,701 \\
    $I_\text{COP}$ & 64.0 & 42.7 \\
    Measured GFLOP/s on TPU V3 & 118,524 & 118,062 \\
    Measured GFLOP/s on TPU V4 & 251,166 & 172,035 \\
    \bottomrule
  \end{tabular}
\end{table}

Table \ref{t:dataset} summarizes the dimensions and kernel properties for the two benchmarks. The memory arithmetic intensity $I_\text{MEM}$ is reported by the TPU profiler, and the instruction throughput intensity $I_\text{COP}$ is manually derived. The following show how we derive $C$ (COPs per dot-product) for each dataset.

\paragraph{Glove} The Glove dataset uses the cosine distance, which yields same search results as MIPS. As described in Appendix \ref{ap:detail}, when the database size is not power-of-2, we pay an extra $C$ in the inner loop. Therefore the total $C$ used for the Glove benchmark are
\begin{itemize}
    \item 3 $C$ by PartialReduce, and
    \item 1 $C$ by non power-of-2 database masking.
\end{itemize}

We pre-process the Glove dataset by padding the dimension from 100 to 128 to avoid one $C$. We are not bottleneck on memory bandwidth so the padding is a good trade-off for better performance.

\paragraph{Sift} The Sift dataset uses the Euclidean distance, which requires more coefficient-wise operations. In Appendix \ref{ap:l2} we showed that we only need to use one extra $C$ for distance computation. However, there are some other inevitable operations used in the benchmark:

\begin{itemize}
    \item 3 $C$ by PartialReduce,
    \item 1 $C$ by the relaxed Euclidean distance computation,
    \item 1 $C$ by non power-of-2 database masking, and
    \item 1 $C$ by broadcasting $\frac{\|\mathbf{x}\|^2}{2}$.
\end{itemize}
 
Therefore the total number of $C=6$, resulting a performance regression on TPU V4 as seen in Figure \ref{fig:multi_roofline}.

\subsection{Alternative implementation}

\begin{listing}[h]
\begin{minted}[frame=lines]{python}
# qy shape: f32[1024,128], db shape: f32[1048576,128]
# output shapes: f32[1024, 128], i32[1024, 128]
@jax.jit
def mips_baseline(qy, db):
  dists = jax.numpy.einsum('ik,jk->ij', qy, db)
  reshaped = jax.lax.reshape(dists, [1024, 128, 8192])
  return jax.lax.argmax(reshaped, 2, jnp.int32)
\end{minted}
\caption{Baseline implementation without the approx\_max\_k operator}
\label{mips:reshape-baseline}
\end{listing}

A naive implementation of Algorithm \ref{alg:approx:mips} can be composed of Reshape and ArgMax. In this section we show that the performance is not comparable to the dedicated \texttt{approx\_max\_k} operator.

Our experiment setup is as follows: let query be $\mathbf{Q}\in \mathbb{R}^{1024\times 128}$
and database be $\mathbf{X}\in \mathbb{R}^{1048576\times 128}$; we choose the reduction output size as $L=128$, so the algorithm can be written as Listing \ref{mips:reshape-baseline}.

We benchmark the implementations on a single-core TPU V4 instance by 100 times and collect the averaged execution time.
Listing \ref{mips:reshape-baseline} took 24.9ms to compute; in comparison, our proposed new operator used in Listing \ref{mips:jax} only took 2.6ms, which is 9.6x faster.

\end{document}